\documentclass[12pt]{article}
\usepackage{amsmath,amssymb,graphicx}
\batchmode

\newtheorem{lem}{Lemma}
\newtheorem{thm}{Theorem}

\newtheorem{cor}{Corollary}

\newcommand{\pr}{\noindent{\bf Proof}. }
\newcommand{\re}{\noindent{\bf Remark}. }
\newcommand{\res}{\noindent{\bf Remarks}. }

\newcommand{\pa}{\partial}

\newcommand{\De}{\Delta}
\newcommand{\de}{\delta}

\newcommand{\Ga}{\Gamma}

\newcommand{\la}{\lambda}
\newcommand{\Om}{\Omega}

\newcommand{\si}{\sigma}

\newcommand{\cC}{{\cal C}}

\newcommand{\cO}{{\cal O}}

\newcommand{\cH}{{\cal H}}

\newcommand{\cE}{{\cal E}}

\newcommand{\cF}{{\cal F}}
\newcommand{\cK}{{\cal K}}

\newcommand{\cN}{{\cal N}}

\newcommand{\cL}{{\cal L}}

\newcommand{\bbR}{{\mathbb{R}}}

\newcommand{\bbC}{{\mathbb{C}}}

\newcommand{\st}{\stackrel{}}

\begin{document}
  
\title{Markov quantum fields on a manifold}
\author{ 
J. Dimock\thanks{Research supported by NSF Grant PHY0070905}
\thanks{dimock@acsu.buffalo.edu}\\
Dept. of Mathematics \\
SUNY at Buffalo \\
Buffalo, NY 14260 }
\maketitle

\begin{abstract}
We study scalar quantum field theory on a compact manifold.  
The free theory  is defined  in  terms of functional integrals. 
For  positive mass it is shown  to have the  Markov property  in the sense of Nelson.
This property is used to establish  a reflection positivity result when the manifold  
has a reflection symmetry.
In  dimension d=2  we use the Markov property to establish a sewing operation for
manifolds with boundary circles.
Also in d=2   the Markov property  is proved   for interacting  fields.   
\end{abstract}

\newpage

\section{Introduction}

We consider a  Riemannian manifold   $(M, g)$ consisting of a  oriented 
compact connected   manifold    $M$ of dimension $d$ and a positive definite
metric $g$.  The  natural inner product  on functions is
 \begin{equation}  
<u,v>=\int \bar  u  v    d \tau  =  \int \bar  u(x)  v(x)  \sqrt{\det g(x)} dx    
\end{equation}
where   $d \tau $ is the Riemannian  volume element and 
 the second expression refers to  local coordinates.
The Laplacian $\De $ can be  defined by the
quadratic form 
\begin{equation}
<u,(-\De)u>=\int  | d u|^2    d \tau  =  \int  g^{\mu \nu}(x) 
  \frac{\pa  \bar u}{\pa x_{\mu}}(x)     \frac{\pa  u}{\pa x_{\nu}}(x)   \sqrt{\det g(x)} dx    
\end{equation}
As is well-known $ -\De$ defines a self adjoint operator in  $\cL^2(M, d \tau )$
with non-negative discrete spectrum and an isolated simple eigenvalue 
at zero  and   with eigenspace the constants.

We want to study
the  free  scalar field of mass  $m  \geq 0 $   on $(M, g)$. For  $m>0$  
this is a family of Gaussian random variables  $\phi(f)=<\phi,f>$ indexed  by smooth real 
 functions $f$ on 
$M$.  The fields $\phi(f)$  are defined to have    mean zero and  covariance $(-\De + m^2)^{-1}$. 
 If $ \mu$ is the underlying measure   we have  the 
characteristic function 
\begin{equation}
\int e^{i\phi(f)}  d \mu    = e ^{-\frac12  <f, (-\De + m)^{-1} f>}
\end{equation}
from which one can generate the correlation functions.

For  $m=0$ 
the  Laplacian is only invertible on the orthogonal complement of the constants
and we restrict the test functions $f$ to lie in this subspace, i.e.  $\int f\ d \tau = 0$.  For 
 $m=0$ and $d=2$,
 metrics  which are equivalent by a local rescaling give rise to  the same fields
\footnote{
For smooth $\lambda >0$ we have  $\De_{\la g} =  \la^{-1} \De_g$ and   hence  
 $<\la^{-1}f, \De_{\la g}^{-1} \la^{-1}f>_{\la g}=<f,\De_{g}^{-1}f>_{g}$.  Thus 
    $\phi_{\la g}( \la^{-1}f)$ and $\phi_{g}(f)$ have the same characteristic
function and are equivalent.},
and we have  a conformal field theory.

In this paper we show that  for  $m>0$  the fields $\phi(f)$  satisfy a Markov property in  the 
sense of Nelson \cite{Nel73a},\cite{Nel73b},\cite{Nel73c}.   Nelson originally developed this
concept  for Euclidean  quantum fields in  $\bbR^n$, and we show that his treatment can  also 
  be carried out on  manifolds. 
We also work out some applications,  generally for  $m>0$ and sometimes by  
limits for  $m=0$. 
We show that  functional integrals can be written as 
inner products of  states  localized
on $d-1$ dimensional submanifolds.
If the manifold has a reflection symmetry this leads to a   
 reflection positivity result and an enhanced Hilbert space structure.
In $d=2$ another application  is 
the establishment of a sewing  property for manifolds with boundary 
circles.  Operations of this type are widely used in conformal field theory and
string theory.
Finally we  obtain the Markov property for interacting  fields  in $d=2$.

\section{Sobolev spaces}

We begin with  some preliminary  definitions. (See for example  \cite{Tay96}).
 Let $H^{\pm 1}(M)$ be the 
usual real  Sobolev spaces consisting of those distributions on   $M$ which when expressed
in  local coordinates are in the spaces   $H^{\pm 1}(\bbR^d)$.
These have no particular norm, but we give an alternate definition
which supplies a norm and an inner product.   
The  spaces     $H^{\pm 1}(M)$ can be identified as 
 completion of   $\cC^{\infty}(M)$ in the norm
\begin{equation}  \label{one}
\|u \|^2_{\pm 1} = <u, (-\De +m^2)^{\pm 1} u >
\end{equation}
for any  $m>0$.  
These are real Hilbert spaces and we have  $H^{1}(M) \subset  \cL^2(M,d \tau)
\subset  H^{-1}(M)$.
   We have also
$|<u,v>|  \leq  \|u\|_{1} \|v\|_{-1} $ so the inner product extends by  limits to 
 a bilinear   pairing of   $H^{1},H^{-1}$.  These spaces 
are dual with respect to this pairing.   Also $-\De +m^2$ is unitary from 
$H^{1}$ to   $H^{-1}$.

For any closed subset  $A  \subset M$
define a closed subspace 
\begin{equation}
H^{-1}_A(M)  =  \{ u \in  H^{-1}(M): \textrm{supp}\ u \subset A\}
\end{equation}
Also for   $\Om  \subset  M$ open let  $ H^1_0(\Om)$ be the closure of $\cC^{\infty}_0(\Om)$
in   $H^1(M) $

Now let     $\Om$ be  open set and consider the 
disjoint unions 
\begin{equation}
 M =  \Om^c \cup \Om \ \ \ \ \ \ M  = ( \textrm{ext} \Om)  \cup \bar \Om  
\  \ \ \ \ \ \ \ M  =  ( \textrm{ext} \Om)   \cup   \pa   \Om  \cup \Om
\end{equation}
For each of these we have an associated decomposition of  $H^{-1}(M)$:

\begin{lem} For open  $\Om \subset M$
\begin{eqnarray} 
 H^{-1}(M)& =  & H^{-1}_{\Om^c} (M) \ \oplus \ (-\De  +m^2) H^1_0(\Om)\\
H^{-1}(M)& = &  (-\De  +m^2) H^1_0(  \textrm{ext}\ \Om) \ \oplus \ H^{-1}_{\bar \Om} (M) \\
H^{-1}(M)& =&   \label{decomp1}
 (-\De  +m^2) H^1_0( \textrm{ext}\ \Om )  \ \oplus \   H^{-1}_{\pa  \Om} (M) \ 
\oplus \ (-\De  +m^2) H^1_0(\Om)
\end{eqnarray}
\end{lem}

\pr  It is straightforward to show that  the orthogonal complement of 
 $H^1_0(\Om)$  in the dual space  $ H^{-1} (M)$ is   $  H^{-1}_{\Om^c} (M)$. 
The dual relation  is  that the orthogonal complement of   $  H^{-1}_{\Om^c} (M)$
in  $H^{1}(M)$ is    $H^1_0(\Om)$.  To find the orthogonal
complement  of   $  H^{-1}_{\Om^c} (M)$  in     $ H^{-1} (M)$  we apply the unitary operator 
$-\De  +m^2$  and get 
$ (-\De  +m^2) H^1_0(\Om) $.  This   gives the first result.

 For the  second result replace  $\Om$ by  $ \textrm{ext} \Om$.

For the third result replace  $\Om$ by   $(\pa \Om)^c$ and obtain
 \begin{equation}
H^{-1}(M) =   H^{-1}_{\pa \Om} (M) \ \oplus \ (-\De  +m^2) H^1_0((\pa \Om)^c)
\end{equation}

The result now follows from
\begin{equation}
 H^1_0((\pa \Om)^c)=   H^1_0(\Om)  \oplus  H^1_0( \textrm{ext} \Om)
\end{equation}
\bigskip

\re  
Applying  the unitary  $(-\De +m^2)^{-1}$ to the decomposition (\ref{decomp1}) of $H^{-1}(M)$
we get a decomposition  of $H^1(M)$
which is  
\begin{equation}  \label{decomp2}
H^1(M) =
 H^1_0( \textrm{ext} \Om)  \ \oplus \  (-\De  +m^2) ^{-1}  H^{-1}_{\pa  \Om} (M) \ 
\oplus  H^1_0(\Om)
\end{equation}
This says that any  element of $H^1(M)$ can be uniquely written  as the  sum of
 a function which satisfies $(-\De +m^2)u = 0$   on $(\pa \Om)^c=  \Om  \cup  \textrm{ext} \Om$
 and a  function which vanishes on $\pa \Om$.
\bigskip
 
 By comparing the various decompositions in the lemma we also have 
corresponding to  $\bar \Om =  \pa \Om  \cup  \Om$ and   $\Om^c = ( \textrm{ext} \Om) \cup  \pa   \Om $
the decompositions:

\begin{cor}
\begin{equation}  \label{more}
\begin{split}
 H^{-1}_{\bar \Om} (M) =& \  H^{-1}_{\pa  \Om} (M) \ \oplus \ (-\De  +m^2) H^1_0(\Om) \\
  H^{-1}_{\Om^c} (M) =& (-\De  +m^2) H^1_0( \textrm{ext}\ \Om)  \  \oplus  H^{-1}_{\pa  \Om} (M)  \\
\end{split}
\end{equation}
\end{cor}
\bigskip

Now  for  $A \subset  M$ let    $e_{A}$ be the orthogonal projection onto  $H^{-1}_A(M)$.
The following pre-Markov property  is basic to our treatment.

\begin{lem}
 For open $\Om \subset  M$
\begin{enumerate}
\item    If   $u \in   H^{-1}_{\bar  \Om}(M)$  then  $ e_{\st{\Om^c}}  u=   e_{\st{\pa  \Om}} u$
\item    $e_{\st{\Om^c}}  e_{\st{\bar  \Om}}= e_{\st{\pa  \Om}} $
\end{enumerate}
\end{lem}
\bigskip

\pr  The two statements are equivalent. With respect to the decomposition   (\ref{decomp1})    
we have 
\begin{equation}
 e_{\st{\Om^c}}=\left( \begin{array}{ccc}1&0&0\\0&1&0\\0&0&0\\\end{array}\right) \ \ \ 
  e_{\st{\bar  \Om}}=\left( \begin{array}{ccc}0&0&0\\0&1&0\\0&0&1\\\end{array}\right) \ \ \  
e_{\st{\pa \Om}}=\left( \begin{array}{ccc}0&0&0\\0&1&0\\0&0&0\\\end{array}\right)
\end{equation}
and hence   $e_{\st{\Om^c}}  e_{\st{\bar  \Om}}= e_{\st{\pa  \Om}} $.
\bigskip

\re 
 If    $u \in   H^{-1}_{ \Om^c}(M)$ and  $ v \in   H^{-1}_{\bar  \Om }(M)$ 
then   
\begin{equation}  \label{collapse}
(u,v)_{-1} = ( e_{\st{\Om^c}}u , e_{\st{\bar  \Om}}v)_{-1} = (u, e_{\st{\pa  \Om}}v)_{-1} 
= ( e_{\st{\pa  \Om}}u, e_{\st{\pa  \Om}}v)_{-1} 
\end{equation}
which reduces the inner product to the boundary.  We can use this to obtain 
a sufficient condition for   $H^{-1}_{\pa \Om}(M)$ to be nontrivial.  (The 
condition is not necessary.)

\begin{cor}
If  $ \Om ,\textrm{ext }\Om \neq  \emptyset$ then   $ H^{-1}_{\pa \Om}(M) \neq \{0\}$
\end{cor}

\pr  
The space  $ H^{-1}_{\pa \Om}(M)$  has a meaning independent of any norm.  It suffices
to show that it is non-trivial as a subspace of   $H^{-1}(M)$  with the norm (\ref{one})
and $m^2$ small.

 Let  $u \in \cC^{\infty}_0( \textrm{ext }\Om)$ and  $ v \in \cC^{\infty}_0(\Om )$  
be positive functions.  We will  show that   $ e_{\st{\pa  \Om}}u \neq 0$  and  $ e_{\st{\pa  \Om}}v \neq 0$.
By  (\ref{collapse}) it suffices to show that  $(u,v)_{-1}   \neq  0$. 
 Let $\psi_0  = 1/\sqrt{ \textrm{Vol} ( M)}$ be the lowest eigenfunction
 of  $-\De$  on  $ \cL^2(M,d \tau)$ .
Then  $u_0 = <u,\psi_0>$ and $v_0 =<v,\psi_0>$ are nonzero.   As  $m \searrow 0$ we have that 
\begin{equation}
(u,v)_{-1} = <u, (-\De +m^2)^{-1} v >  =  u_0v_0 m^{-2}  +  \cO(1)
\end{equation}
Thus $(u,v)_{-1}  \neq 0$  
 for $m^2$ small.

\section{Markov property}

We use these results to establish the Markov property for 
our $m>0$ field theory following Nelson   \cite{Nel73c}.
First extend the class of test functions from  $\cC^{\infty}(M)$  to   $H^{-1}(M)$ so that 
now   $\phi(f)$ is a family of Gaussian random variables indexed by 
$f \in H^{-1}(M)$  with  covariance given by the $H^{-1}(M)$ inner product.
The underlying measure space $(Q,\cO,\mu)$ consists of a set  $Q$,
a $\si$-algebra  of measurable subsets  $\cO$ generated by the  $\phi(f)$, and a measure   $\mu$.
Polynomials  in  $\phi(f)$ are dense in $\cL^2(Q,\cO, d \mu)$. 
We also need
Wick monomials $:\phi(f_1) \dots \phi(f_n):_{\st{(-\De + m^2)^{-1}}}$
defined as the projection in  $\cL^2(Q,\cO, d \mu)$  of  $\phi(f_1) \dots \phi(f_n)$
onto the orthogonal complement of polynomials of degree  $n-1$.
These are polynomials of degree  $n$ and for example
\begin{equation}
   :\phi(f) \phi(g):_{\st{(-\De + m^2)^{-1}}}\ =\  \phi(f) \phi(g)
-  <f,(-\De + m^2)^{-1}g> 
\end{equation}

Let us  recall the well-known connection between the Gaussian
processes and Fock space. 
Let   $\cF(H^{-1}_{\bbC})$  be
the Fock space over the complexification $H_{\bbC}^{-1}(M)$,
that is   the infinite direct sum of $n$-fold symmetric
tensor products of the  $H^{-1}_{\bbC}(M)$.
Then there  is an unitary identification  of  (complex) 
 $ \cL^2( Q, \cO, d \mu)$  with   $\cF(H^{-1}_{\bbC})$
determined by  
\begin{equation}
:\phi(f_1) \dots \phi(f_n):_{\st{(-\De + m^2)^{-1}}}\
\ \    \leftrightarrow  \ \ \ \  \sqrt{n!}\ \textrm{Sym} (f_1 \otimes \dots  \otimes  f_n) 
\end{equation}
Any  contraction $T$   on $H^{-1}_{\bbC}(M)$  (linear operator with $\|T\| \leq 1$)
  induces a    contraction  $\Ga(T)$ on the Fock space
by sending   $\textrm{Sym} (f_1 \otimes \dots  \otimes  f_n) $
to  $\textrm{Sym} (Tf_1 \otimes \dots  \otimes  Tf_n) $.
This determines  a contraction on 
$ \cL^2( Q, \cO, d \mu)$ also denoted  $\Ga(T)$.  We have  $\Ga(T) \Ga(S) = \Ga(TS)$.

Now for  closed    $A \subset  M$ let   $\cO_A$ be the smallest  subalgebra of  $\cO$
such that the functions  $\{ \phi(f):   supp \ f \subset A\}$ are measurable. Also let
$\cE_A F = \cE\{ F| \cO_A\}$ 
be the conditional expectation of a function  $F$ with respect to  $\cO_A$.
 Then  $\cE_A$ is an orthogonal projection on  $\cL^2(Q,\cO, d \mu)$
with range    $\cL^2(Q,\cO_A, d \mu)$, the  $\cO_A$ measurable $\cL^2$-functions.

The conditional expectations are related to  the projections
in Sobolev space by 
\begin{equation}  \label{proj}
\cE_A  =  \Ga (e_A) 
\end{equation}
For the proof   see Simon \cite{Si75}.  This leads to

\begin{thm}  (the Markov property)  For open   $\Om \subset M$
\begin{enumerate}
\item    If   $F \in  \cL^2(Q,\cO_{\bar \Om}, d \mu)$  then  $ \cE_{\Om^c}F=   \cE_{\pa \Om}F$
\item    $\cE_{\Om^c}\cE_{\bar\Om}=  \cE_{\pa \Om} $
\end{enumerate}
\end{thm}
\bigskip

\pr  The two statements are equivalent.  The  second
 follows from   $e_{\st{\Om^c}}  e_{\st{\bar  \Om}}= e_{\st{\pa  \Om}} $ and  (\ref{proj}) for we have  
\begin{equation}
\cE_{\Om^c}\cE_{\bar\Om} =  \Ga( e_{\st{\Om^c}})\Ga (e_{\st{\bar  \Om}} ) 
=  \Ga( e_{\st{\Om^c}}e_{\st{\bar \Om}} ) 
=  \Ga(e_{\st{\pa \Om}}) 
=  \cE_{\pa \Om}
\end{equation}
\bigskip

\re 
 Now suppose that  $F$ is $\cO_{ \Om^c}$ measurable and 
$G$ is   $\cO_{\bar  \Om}$ measurable.   Then by $\cE_{\Om^c}\cE_{\bar\Om}=  \cE_{\pa \Om}$ 
we have 
\begin{equation}   \label{collapse3}
\int  \bar  F G d \mu  =\int  \overline{  \cE_{\Om^c}  F}(\cE_{\bar\Om}  G) d \mu 
=\int   \bar F (\cE_{\st{\pa  \Om}}G)  
=  \int \overline{\cE_{\st{\pa  \Om}}F}( \cE_{\st{\pa  \Om}}G)   d \mu 
\end{equation}
This says  that the conditional expectation  $ \cE_{\st{\pa  \Om}}$
maps    $\cO_{\bar \Om}$   measurable functions   and   $\cO_{\Om^c}$ measurable functions
to  $\cO_{\pa  \Om}$  measurable functions in such a way 
that the functional integral is evaluated as the inner product in the boundary Hilbert space
$ \cL^2(Q, \cO_{ \pa \Om}, d \mu)$
We exploit this identity in the next two sections.

\section{Reflection positivity}

As a first application we show that  if the 
manifold has a reflection symmetry  then  the functional integrals
have a more elementary  Hilbert space structure.
We assume that our $d$-dimensional manifold  $M$ has a $d-1$ dimensional
submanifold  $B$ which divides the manifold in two identical parts.
That is we have the disjoint union  
\begin{equation}
M  =  \Om_{-}  \cup    B\   \cup    \Om_+
\end{equation}
where   $\Om_{\pm}$ are open and  $\pa \Om_{\pm}  =  B$.
Further we assume 
there is an  isometric involution    $\theta$  on  $M$ so that   $\theta  \Om_{\pm} = \Om_ {\mp}$
and    $\theta  B = B$.   For  $d=2$ this is the structure of a Schottky double.
As an example in  $d$ dimensions we could take  $M$ to be  the sphere 
   $\{ x \in \bbR^{d+1}: x_0^2 + \dots + x_d^2=1\} $,
   take  $B = \{ x_0 =0\}$ and     $\Om_{\pm}  =  \{  \pm x_0 >0\} $,
and  let   $\theta$ be the  reflection in  $x_0  \to - x_0$.

As a diffeomorphism $\theta$ defines a map   $\theta_*$ on
$\cC^{\infty}(M)$ by   $\theta_* u = u \circ  \theta^{-1} $  
which  extends to a bounded operator on   $H^{\pm 1}(M)$ or  $\cL^2(M)$.
Since   $\theta$ is an isometry $\theta_*$ is unitary  on these spaces
and preserves the  $H^1,H^{-1}$ pairing.  Since
$\theta^2 =1 $ we have   $(\theta_*)^2 =1$. 

\newpage

\begin{lem} Let  $u \in H^{-1}_{B}(M)$.
\begin{enumerate}
\item  $<u,f> =0$ for any smooth function vanishing on  $B$.
\item  $\theta_*  u = u$.
\end{enumerate}
\end{lem}

\pr
 By choosing local coordinates we  
reduce (1.)  to the following statement.
Let $u \in   H^{-1}_{B_0}(\bbR^d)$ 
where  $B_0 =\{x \in \bbR^d:x_d =0\}$
and  let   $f \in \cC^{\infty}_0(\bbR^d)$ vanish on $B_0$. 
Then $<u,f>=0 $.  A distribution with support in $B_0$ 
is a finite  sum of derivatives of  delta functions:   $u = \sum_j  h_j \otimes \de^{(j)}_{B_0}$.
The condition  $f\in H^{-1}(\bbR^d)$ rules out $j \geq 1$ as can be 
seen by looking at the Fourier transform.  Thus $u = h \otimes \de_{B_0}$   
and the result follows.

For (2.)   we must show that   $<\theta_*u  -u,f>=0$  for smooth $f$ 
or equivalently that $<u,f - \theta_*f> =0$. 
Since   $f - \theta_* f $ vanishes on   $B$ this follows from part one.
This completes the proof.
\bigskip

Now let   $\Theta =\Ga(\theta_*)$  be the induced
reflection on  $\cL^2(Q, \cO, d  \mu)$.   This is unitary since
$\theta_*$ is unitary and we also have 
\begin{equation}
\Theta  (\ \phi(f_1) \dots \phi(f_n)\ ) =  \phi(\theta_*f_1) \dots \phi(\theta_*f_n)
\end{equation}

\begin{thm} (Reflection Positivity,  $m > 0$ )
For     $F  \in  \cL^2(Q, \cO_{\bar  \Om_+}, d  \mu)$
 \begin{equation}  \label{action}
\int \overline{ \Theta(F)} F  d \mu  \geq 0 
\end{equation}

\end{thm}
\bigskip

\res   The positivity  is also known as Osterwalder-Schrader  positivity. 
A similar result was previously obtained by  
 De Angelis, de Falco,  Di Genova \cite{DDD86} by other methods.
 The proof below follows Nelson  \cite{Nel73c}.
\bigskip

\pr
For any closed set  $A$ we have 
 $\theta_* H^{-1}_A  =  H^{-1}_{\theta A}$
and hence   $\theta_* e_{\st{A}}  = e_{\st{\theta A}}  \theta_*$.
It follows that 
\begin{equation}
\Theta \cE_{A} = \Ga( \theta_*)  \Ga( e_{\st{A}})=  \Ga( e_{\st{\theta A}} )  \Ga( \theta_*) =  \cE_{\theta A}  \Theta
\end{equation}
In particular we have   $\Theta \cE_{  \bar  \Om_+} =  \cE_{\Om_+^c}  \Theta $
and    $\Theta \cE_{B} =  \cE_{B}  \Theta $.

 The result  now  follows by the calculation 
\begin{equation}  \label{calc}
\int \overline {(\Theta F)} F d \mu 
 = \int  \overline{  \cE_{B}(\Theta F) } \   \cE_{B}F \ d \mu 
= \int | \cE_{B}( F) |^2 d \mu  \geq 0
\end{equation}
Here in the first step we have used   $\Theta \cE_{  \bar  \Om_+} =  \cE_{\Om_+^c}  \Theta $
to conclude that   $\Theta F$ is  $\cO_{\Om_+^c}$ measurable and 
then  (\ref{collapse3}) to reduce the 
calculation to $B$.   For the  second 
step we note that the lemma says   $ \theta_* e_{\st{B}} = e_{\st{B}}$  and so    
 $\Theta \cE_{B} =  \cE_{B}$.  Hence   $\cE_{B} \Theta =  \cE_{B}  $
to complete the proof.
\bigskip

Next we consider the case $m=0$ as defined in the introduction.  Let 
$\mu_{0}$ denote the measure and  again  define $\Theta$  so that  (\ref{action}) holds.  We take 
a smaller class of functions $F$ but otherwise have the same result.

\begin{cor}   (Reflection Positivity,  m=0 )  \label{m=0}
 Let    $F$  be a polynomial in the fields 
$  \phi(f)$ with $  f \in \cC^{\infty}_0(\Om_+)$ and $ \int f d \tau =0 $.
Then    
\begin{equation}
\int \overline{ \Theta(F)} F  d \mu_0  \geq 0 
\end{equation}
\end{cor}

\pr  If $f,g$ satisfy  $\int f d \tau =0$
  then  $<f, (-\De)^{-1} g>  =  \lim_{m \to 0}   <f, (-\De+ m^2)^{-1} g>$.
 Gaussian integrals of  polynomials  can be explicitly evaluated as sums of products of 
these expressions.  Hence  if $P$ is  any polynomial with these test functions 
 and  $\mu_m$   the  massive measure then
   $ \int  P  d \mu_0   =   \lim_{m \to 0}  \int  P  d \mu_m$.  In particular 
\begin{equation}
 \int \overline{ \Theta(F)} F  d \mu_0   =   \lim_{m \to 0} 
 \int \overline{ \Theta(F)} F  d \mu_m   
\end{equation}
The result now follows from the previous theorem.
\bigskip

\res  Returning to the case  $m>0$ one can now 
 define an inner product on  $\cO_{\bar \Om^+}$ measurable functions $F,G$
 by 
\begin{equation}
<F,G> = \int \overline{ \Theta(F)} G  d \mu 
\end{equation}
Then   $<F,F>\ \geq 0$ and if we divide out  the null vectors $\cN  = \{ F: < F, F>=0 \}$
 we get something positive definite and hence a pre-Hilbert space.
 We call the  Hilbert space  completion   $\cK$:
\begin{equation}
\cK=  \overline{ \cL^{2} (Q,\cO_{\bar \Om^+}, d \mu) /\cN }
\end{equation}
A similar construction works for   $m=0$.
 
Now we are in a position to define operators on  $\cK$ 
from certain operators  on the  $\cL^2$ space.   For details 
on such constructions  and related positivity results in conformal field 
theory see 
 \cite{FFK89},  \cite{Gaw99},  \cite{JKL89}.

\section{Sewing}  \label{sew}

Now restrict to $d=2$  and suppose that we have a Riemann surface  $(M_1,g_1)$ with  
a boundary circle $C_1$.  Further  suppose that  
the metric is flat on a neighborhood of the boundary.
This means that there is a local coordinate   $z$ in  which the circle is $|z| =1 $
the metric has the form   $|z|^{-2} dz d  \bar z$ for   $|z| > 1$.
If we allow ourselves local rescalings of  of the metric  $g \to \la g$
this is not a restrictive condition.  These rescalings
are permitted if $m=0$.  Even if   $m>0$ the effect
of such a transformation  would be to change to a variable
mass, and this would not spoil our results.

  We want to define a mapping from an algebra of   fields on  $M_1$ to  
states  on the boundary  $C_1$. 
We have already noted that for  a manifold without boundary  the conditional 
expectation  serves this function,  so we proceed by closing $M_1$.
That is   we cap off the circle in some standard fashion  to get 
a compact manifold   $(\tilde M_1, \tilde g_1)$ without boundary, also flat 
in a neighborhood of   $C_1$.  
Then for  $m>0$ we have Gaussian  fields   $\{ \phi_1(f): f \in H^{-1}(\tilde M_1) \}$  
on a measure space  $( Q_1,  \cO_{1} ,   \mu_1)$.
As the  boundary Hilbert 
space we take the  $\cL^2$ functions measurable with respect to  $\cO_{1,C_1}$: 
\begin{equation}
\cH_{C_1} \equiv  \cL^2(  Q_1,  \cO_{1,C_1} ,  d \mu_1)
\end{equation}
Then  we define   
\begin{equation}
A_{C_1,M_1}: \cL^2(  Q_1, \cO_{M_1},  d \mu_1) \to  \cH_{C_1}
\end{equation}
as the restriction of the conditional expectation in  $\tilde M_1$
\begin{equation}
A_{C_1,M_1}  =   \cE^{\tilde M_1}_{C_1}  
\end{equation}
We further restrict the domain to the algebra of polynomials in 
$ \{  \phi_1(f):  f \in H^{-1}_{M_1}(\tilde M_1)  \} $.

Suppose also there is  a second such Riemann surface   $(M_2, g_2)$ with boundary circle 
$C_2 $ and  a local coordinate  in which the circle is $|w| =1 $
the metric has the form   $|w|^{-2} dw d  \bar w$ for  $|w| > 1$. We cap off  $M_2$
to form a manifold without boundary
$(\tilde   M_2, \tilde g_2)$.  Then we have fields
 $\{ \phi_2(f): f \in H^{-1}(\tilde M_2) \}$
on a measure space  $( Q_2,  \cO_{2} ,   \mu_2)$, and a 
operator    $A_{C_2,M_2}  =   \cE^{\tilde M_2}_{C_2}  $.

The two manifolds  $M_1,M_2$ can be joined together  by 
identifying points in a neighborhood of $C_1$ in  $\tilde M_1$
with points in a neighborhood   of $C_2$ in  $\tilde M_2$
when the coordinates satisfy  $z=1/w$. Then  $C_1$ and $C_2$ 
are identified by an orientation reversing map.  On the overlap we have two coordinates
and two metrics,  but  the metrics agree since the coordinate
change  $z=1/w$ takes $|z|^{-2} dz d  \bar z$ to   $|w|^{-2} dw d  \bar w$.
 Thus we get a compact 
Riemann  surface   $(M,g)$ which is flat in a neighborhood 
of a  circle   $C$.   (see figure 1, and 
see   \cite{Hua97} for more  details on this construction).    There is   an  isometric 
mapping  $j_1$  from   a  neighborhood of   $M_1$ in $\tilde M_1$
into   $M $ which   takes $C_1$  to $C$.  The image of  $M_1$  in $M$ will also be called  $M_1$.
 Similarly  we have an isometric mapping    $j_2$ from
a neighborhood of    $M_2$   in  $\tilde M_2$
to   $M$  which takes   $C_2$ to  $C$.

\begin{figure}[p] 
\includegraphics*[0in,5.5in][4.25in,11in]{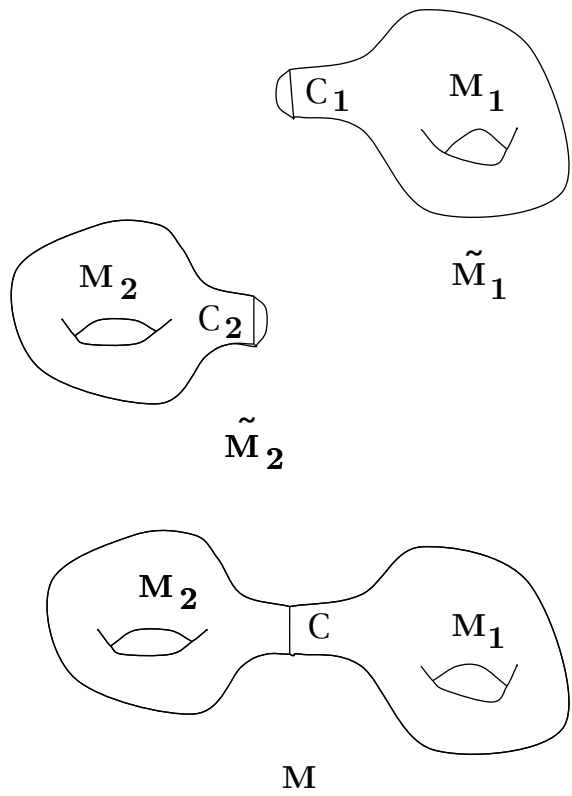}
\caption{$M_1,M_2$ are manifolds with boundary circles $C_1,C_2$.   They are capped off
to form $\tilde M_1, \tilde M_2$.  They are sewn together to form the manifold $M$ without 
boundary}   \label{sewing}
\end{figure}

On the new manifold $M$ we have Gaussian   fields 
   $\{ \phi(f): f \in H^{-1}( M) \}$  on a measure space  $( Q,  \cO ,   \mu)$.
We also have an identification between fields  on 
$M_1$ in $\tilde M_1$ and fields  on $M_1$ in $M$.  To see this 
first note that the  isometry   $j_1$ induces a map 
$  j_{1,*}$ from  distributions on  $\tilde M_1$ with support in  $M_1$ to 
 distributions on  $M$ with support in  $M_1$.  This map preserves 
Sobolev spaces and so  
\begin{equation}
 j_{1,*}:   H^{-1}_{M_1} (\tilde M_1)\to   H^{-1}_{M_1} ( M)
\end{equation}
However with our nonlocal norms  (\ref{one}) this is not unitary.
There is an induced map on Fock space subspaces:
\begin{equation}
J_1  \equiv  \Ga( j_{1,*}):   \cF( H^{-1}_{M_1} (\tilde M_1))
\to  \cF( H^{-1}_{M_1} ( M))
\end{equation}
Since $ j_{1,*}$ is not  a contraction  $J_1$ is  unbounded.
We take as the domain elements with a finite number of entries.
 We can also regard $J_1$ as a map of the corresponding  $\cL^2$ subspaces
\begin{equation}
J_1: \cL^2(  Q_1,  \cO_{1,M_1} ,  d \mu_1) \to  \cL^2(  Q,  \cO_{M_1} ,  d \mu) 
\end{equation}
with domain the polynomials. 
We note also that  $J_1$ maps $\cH_{C_1}$
to $\cH_{C} \equiv  \cL^2(  Q,  \cO_{C} ,  d \mu)$.
There is a similar map $J_2$.

Our goal is to sew together the operators  $A_{C_1,M_1}$ and $A_{C_2,M_2}$
and obtain  an managable functional integral on  the new manifold   $M$.
The recipe is as follows.  Starting with 
 polynomials     $F,G$  on  $M_1,M_2$
we  propagate them to the circles   $C_1,C_2$  by forming  
$A_{C_1,M_1}F$ and  $A_{C_2,M_2}G$.  Then we map to the circle   $C$
forming $J_1A_{C_1,M_1}F$ and  $J_2A_{C_2,M_2}G$  in   $\cH_C $.  Finally  we take 
the inner product  in $\cH_C$.
Thus we define 
\begin{equation}
(A_{C_1,M_1}F, \ A_{C_2,M_2}G)
=  \int \overline { (J_1A_{C_1,M_1}F)}(J_2A_{C_2,M_2}G) d \mu
\end{equation}

\begin{thm} (Sewing,  $m>0$) Let  $F$ be a polynomial in
  $\{ \phi_1(f):  f \in H_{M_1}^{-1} (\tilde M_1)\}$  and let $G$ be  
a polynomial in  $\{ \phi_2(f):  f \in H_{M_2}^{-1} (\tilde M_2)\}$.   Then
\begin{equation}
(A_{C_1,M_1}F, \ A_{C_2,M_2}G)
=  \int  \overline  { ( J_1F)}(J_2G) d \mu
\end{equation}
\end{thm}
\bigskip

\re   Thus  sewing involves the identification operators   $J_1,J_2$ from  $M_1,M_2$ to  $M$.
These can be understood as a change in Wick ordering.  We have   
\begin{equation}
J_1 \left(: \phi_1(f_1) \dots  \phi_1(f_n) :_{(-\De_{\tilde M_1}+m^2)^{-1} } \right)
=: \phi(j_{1,*}f_1) \dots  \phi(j_{1,*}f_n) :_{(-\De_{M}+m^2)^{-1} }
\end{equation}
\bigskip

\pr  
We have that   $j_{1,*}$ maps   $H^{-1}_{M_1}(\tilde  M_1)$ to  
$H^{-1}_{M_1}(M)$.   These spaces have the decompositions (\ref{more})
\begin{equation}
\begin{split}
H^{-1}_{M_1}(\tilde  M_1) =&H^{-1}_{C_1}(\tilde  M_1)  \oplus (- \De_{ \tilde M_1} + m^2)H_0^1(\textrm{int }  M_1) \\
H^{-1}_{M_1}(M)   =& H^{-1}_{C}(M)  \oplus (- \De_{M} + m^2)H_0^1(\textrm{int }  M_1) 
\end{split}
\end{equation}
and since $j_1$  is an isometry  $j_{1,*}$ preserves the decomposition.
The operators   $ e^{\tilde M_1}_{C_1}$ and  $ e^{ M}_{C}$ 
are the projections onto the first factors and so we have  
the identity  on $H^{-1}_{M_1}(\tilde  M_1)$ 
\begin{equation}
j_{1,*}\ e^{\tilde M_1}_{C_1}  =  e^{ M}_{C}\ j_{1,*}  
\end{equation}
It follows that  
\begin{equation}
J_1\ \cE^{\tilde M_1}_{C_1} = \Ga(j_{1,*}) \Ga( e^{\tilde M_1}_{C_1})=
\Ga( e^{ M}_{C})\Ga( j_{1,*})=    \cE^{M}_{C}\ J_1  
\end{equation}
Then we have  
\begin{equation}
\begin{split}
(A_{C_1,M_1}F, \ A_{C_2,M_2}G)
= & \int  \overline{ ( J_1  \cE^{\tilde M_1}_{C_1}F)}(J_2  \cE^{\tilde M_2}_{C_2}G) d \mu \\
= & \int  \overline{ ( \cE^M_{C} J_1  F)}( \cE^M_{C} J_2  G) d \mu \\
= & \int \overline{ (  J_1  F)}(  J_2  G) d \mu \\
\end{split}
\end{equation}
In the last step we use that   $J_1F$ is $\cO_{M_1}$ measurable, that   $J_2G$ is $\cO_{M_2}$
measurable, and  the Markov property via the identity (\ref{collapse3}).
This completes the proof.
\bigskip

\res

\noindent (1.)  We do not attempt a direct sewing result in the case $m=0$.
However one can get something in this direction by restricting 
the class of test functions and taking the limit   $m \to 0$ as in 
Corollary  \ref{m=0}.
\bigskip

\noindent (2.)  Our treatment has featured manifolds with a single boundary 
circle.   However one could as well consider manifolds with many 
boundary circles  $\{C_i\}$. In this case one would consider  operators between 
(algebraic) tensor products of Hilbert spaces   $\cH_{C_i}$ based on the 
boundary circles.   Again one can show a sewing property of 
the type we have presented.  This is essentially  the 
structure discussed by 
 Segal   \cite{Seg89} in his axioms for conformal field theory,  except that we have not accommodated the
possibility of sewing together boundary circles  on the same manifold.  See also   
Gawedski \cite{Gaw99},  Huang  \cite{Hua97}, and Langlands \cite{La02}.

\section{Interacting fields}

We continue to restrict to  $d=2$  and now study interacting fields on a compact Riemann surface  
$(M,g)$. 
For this we may as well assume $m>0$.
We introduce
a potential for  $A \subset  M$  
\begin{equation}
V_{A}(\phi)  =   \int_{A} : P(\phi(x)):_{(-\De +m^2)^{-1}}  \sqrt{ \det g(x)} dx
\end{equation}
Here  $P$ is a lower semi-bounded polynomial.
This not obviously well-defined since it refers to 
products of distributions.   However it turns out that  the  Wick ordering provides sufficient 
regularization and we have 

\begin{lem}  
$V_{A},  e^{-V_{A}}$ are functions  in $ \cL^p(Q,\cO, d \mu)$
for all   $p< \infty$.
\end{lem}

In the plane and with  $A$ compact
 this is a classic result of constructive field theory.
\cite{Nel73c},  \cite{Si75}, \cite{GJ87}.  The proof has been extended to compact subsets of  
 paracompact complete  Riemannian
manifolds by  De Angelis, de Falco,  Di Genova \cite{DDD86}.  Hence it holds
for compact manifolds and 
an interacting field theory  can  be   defined by the measure
 \footnote{ See \cite{Di84} for some results on Lorentzian manifolds}  
\begin{equation}
d \nu  =   \frac{  e^{-V_M}\ d \mu}{ \int  e^{-V_M}  d \mu} 
\end{equation}
 As noted by  Gawedski  \cite {Gaw99} there may be special  choices of  the polynomial $P$
 such that this is a  conformal field theory.

For each measure   $\nu$ we have the
   conditional expectation $\cE_{A}^{\nu}F= 
 \cE^{\nu}\{ F |  \cO_{A}\}$.
This conditional expectation  can be  expressed in terms of the conditional expectation $\cE_{A}$
for  $\mu$ by  
\begin{equation}
\cE_{A}^{\nu} F =   \frac  {\cE_{A}( F e^{- V_{A^c}})}{\cE_{A}( e^{- V_{A^c}})}
\end{equation}
See \cite{Si75} for this identity.
Now   the Markov property for  $\nu$ follows directly from 
the Markov property for $\mu$.  This 
is the following  which  generalizes the 
result of Nelson on the plane  \cite{Nel73c}: 

\begin{thm}  For open  $\Om \subset M$, let  $F$ be   $\cO_{\bar \Om}$ measurable.  Then
\begin{equation}
\cE^{\nu}_{\Om^c} F  =  \cE^{\nu}_{\pa \Om } F
\end{equation}
\end{thm}
\bigskip

\noindent
\textbf{Acknowledgment:}
This work was initiated at the Institute for Advanced Study in Princeton  
whose hospitality I gratefully acknowledge.


\begin{thebibliography}{99}





\bibitem{DDD86} G. De Angelis, D.   de Falco,  G. Di Genova,     
Random fields on Riemannian manifolds: a constructive approach.
Commun. Math. Phys. \textbf{103}, 297-303, (1986)

\bibitem{Di84}  J. Dimock,  $P(\phi)_2$ models with variable coefficients, 
Ann. of Phys. \textbf{154}, 283-307, (1984).



\bibitem{FFK89}  G. Felder, J. Frohlich, J. Keller, On the structure
of unitary conformal field theory, Commun. Math. Phys. \textbf{124}, 417-463, (1989)


\bibitem{Gaw99}  K. Gawedski, Lectures on conformal field theory,  in Quantum fields and strings: a course for 
mathematicians,   P. Deligne, et. al. ,eds, American Mathematical Society, Providence, 1999. 


\bibitem{GJ87}  J. Glimm, A. Jaffe,  Quantum physics, an functional integral
point of view. Springer, New York, 1987.


\bibitem{Hua97}  Y.Z. Huang,   Two  dimensional conformal geometry and vertex 
operator algebras,  Birkhauser, Boston, 1997. 


\bibitem{JKL89} A. Jaffe, S.Klimek, A. Lesniewski, 
   Representations of the Heisenberg algebra on a Riemann 
surface,  Commun. Math. Phys. \textbf{126}, 421-431, (1989)


\bibitem{La02}  R.P. Langlands,     The renormalization fixed point as
a mathematical object,   IAS preprint.


\bibitem{Nel73a} E. Nelson, Construction of quantum fields from Markov fields,
J. Func. Anal. \textbf{12}, 97-112, (1973).

\bibitem{Nel73b} E. Nelson,  The free Markov field,  J. Func. Anal. \textbf{12}, 211-227, (1973).


\bibitem{Nel73c}  E. Nelson, Probability theory and Euclidean field 
theory,
in G. Velo., A. Wightman, eds, {\em  Constructive Quantum field theory}, Springer-Verlag,
New York, 1973.

\bibitem{Seg89}  G. Segal,  Two-dimensional conformal field theories and modular 
functions, IX International Congress on Mathematical Physics, B. Simon, A. Truman,
and I.M. Davies, eds.,22-37, Adam Hilger, 1989.

\bibitem{Si75}  B. Simon,  The $P(\phi)_2$ Euclidean field theory, Princeton University Press,
Princeton, 1974.


\bibitem{Tay96}  M. Taylor,  Partial Differential Equations I,  Springer, New York,  1996.

\end{thebibliography}
\end{document}